# MEASUREMENT TECHNIQUE FOR ELASTIC AND MECHANICAL PROPERTIES OF POLYCRYSTALLINE SILICON-GERMANIUM FILMS USING SURFACE ACOUSTIC WAVES AND PROJECTION MASKS

*Abdelali Bennis, Christina Leinenbach, Carsten Raudzis, Roland Müller-Fiedler, Silvia Kronmüller*

Robert Bosch GmbH, Corporate Research, Stuttgart, Germany

## ABSTRACT

Using Rayleigh surface acoustic waves (SAW), the Young's modulus, the density and the thickness of polycrystalline Silicon-Germanium (SiGe) films deposited on silicon and $SiO_2$ were measured, in excellent agreement with theory. The dispersion curve of the propagating SAW is calculated with a Boundary Element Method (BEM)-Model based on Green's functions.

The propagating SAW is generated with a nanosecond laser in a narrowband scheme projecting stripes from a mask on the surface of the sample. For this purpose a glass mask and a liquid crystal display (LCD) mask are used. The slope of the SAW is then measured using a probe beam setup. From the wavelength of the mask and the frequency of the measured SAW, the dispersion curve is determined point by point.

Fitting the BEM-Model to the measured nonlinear dispersion curve provides several physical parameters simultaneously. In the present work this is demonstrated for the Young's modulus, the density and the thickness of SiGe films.

The results from the narrowband scheme measurement are in excellent agreement with separated measurements of the thickness (profilometer), the density (balance) and the Young's modulus (nanoindenter).

## 1. INTRODUCTION

Polycrystalline silicon-germanium (SiGe) plays an increasing role in micro-electro-mechanical-systems (MEMS) for example as structural material for MEMS integration technology. Compared to polycrystalline silicon as a standard functional MEMS-material, SiGe provides lower deposition temperatures and yields similar physical properties to silicon. In that way it is an alternative functional layer material and can be deposited on top of active IC area.

For the application of SiGe as functional layer accurate values of the elastic and mechanical properties, such as Young's modulus, density and thickness, are needed. Ultrasound techniques using lasers for generating and detecting surface acoustic waves (SAWs), provide a versatile tool to determine simultaneously several film properties.

## 2. TECHNOLOGY

For the investigation of thin film characteristics polycrystalline SiGe layers were deposited by LPCVD-technology. The deposition system, a horizontal type batch reactor which allows favorably simultaneous processing of sixty 6"-wafers. The deposition tool offers temperatures up to 620°C and pressures up to 2 Torr. As precursor gases Silane ($SiH_4$) and germane (10% $GeH_4$ in hydrogen) were used. For in-situ doping of the layers small amounts of phosphine ($PH_3$) or diborane ($B_2H_6$), diluted in $H_2$ can be added. The layers were deposited on monocrystalline 6" silicon wafers as well as on thermally oxidised silicon wafers with 2.5μm $SiO_2$. The pressure was in the range of 200 mTorr. The flow rates of germane and silane were varied to change the composition of the SiGe layers. In this paper the investigation of the elasto-mechanical properties of these LPCVD-SiGe layers are reported with respect to the germanium concentration as well as the kind of substrate from which the film growth started.

## 3. EXPERIMENTAL SETUP

The narrowband scheme generating illuminated periodic patterns on the surface of the sample for the excitation of narrowband SAW can be realized using the transient laser-grating technique [1]. A more precise method consists in using projection masks [2]. In a first experiment a system of several masks was employed to generate a variety of line patterns with periods in the range between 24 and 96μm. For the detection, a broadband laser probe-beam-deflection (PBD) setup was employed [3].

In the present work the use of a more versatile liquid crystal display (LCD) mask is also demonstrated. This mask is fixed in the optical setup and the period of its grating is changed electronically. The potential of both relatively simple setups is demonstrated by nonlinear dispersion measurements in a two-layer system with a few μm thick SiGe film deposited on silicon with a silicon-oxide interface layer. These experiments require neither structuring nor any chemical preparation of the samples. This optical measurement technique is non-destructive and non-contact.

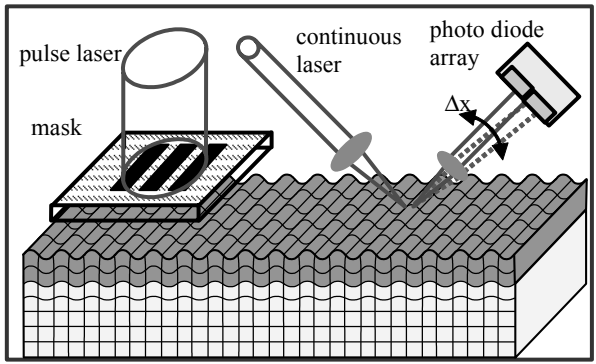

Figure 1: Experimental setup with a glass mask for the excitation of SAW.

In figure 1 the experimental setup is shown. A frequency-doubled Nd:YAG laser with a wavelength of 532 nm, 0.5 mJ pulse energy, and 1.2 ns pulse duration at full width half maximum is used to excite SAW trains with a series of higher harmonics by illuminating selected masks. In the PBD-setup, a continuous wave 100 mW diode-pumped Nd:YAG laser (532 nm) is used to detect the propagating SAWs at a distance of ~5 mm from the source. With this probe method the slope of the transient surface displacement of the propagating SAW is measured.

The set of photomasks (ML&C, Jena) with periods $d$ of 24, 32, 48, 64, and 96 μm is placed at a distance of ~0.1mm from the surface of the samples. The line pattern of the mask is projected at a ratio of 1:1 on this surface to launch the narrowband SAW. The complete set of masks available consists of 15 individual square masks of 10×10 mm$^2$ size, with parallel metal bars of period $d$ and width $h$ and a $h/d$ ratio varying between 0.3 and 0.7. The mask material is quartz and the metal lines consist of chromium with an anti-reflection coating. The number of periods of the mask line patterns varies between 100 for the 96μm mask and 400 for the 24μm mask. This high number of periods ensures a narrowband excitation with a bandwidth of less than 1%.

The amplitude-frequency spectra obtained as fast Fourier transforms (FFTs) from the registered SAW signal contain several narrow peaks, each of which corresponds to a wavelength multiple of the fundamental period of the particular mask applied (Fig. 2). The phase velocity $V_{ph}$ is calculated using the relation $V_{ph} = f \cdot d$, where $f$ are the frequencies of the FFT analysis and $d$ is the wavelength defined by the mask period.

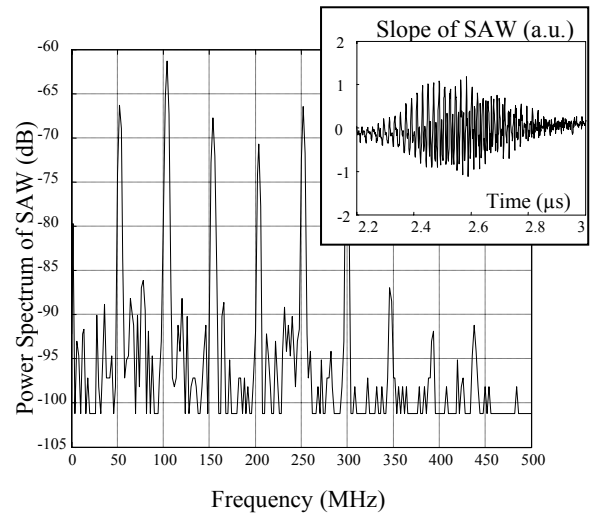

Figure 2: Slope and FFT of the narrow-band SAW from the glass mask measurement for sample 1A (see Table 2).

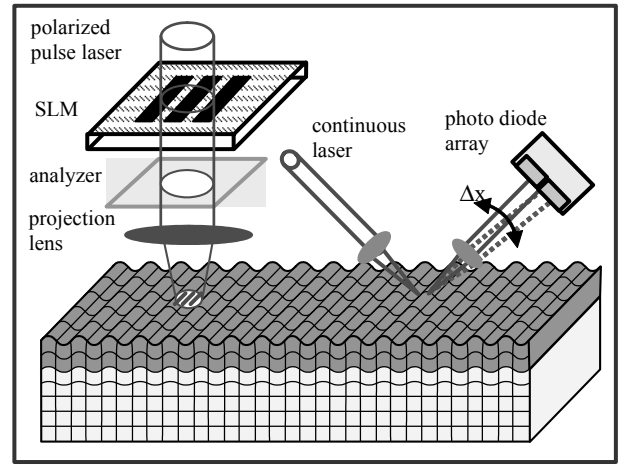

Figure 3: Experimental setup with an SLM mask for the excitation of SAW.

In a second setup a Spatial Light Modulator (SLM) as LCD mask is used instead of the glass mask. The SLM (Holoeye LC 2002) has a pixel pitch of 32μm and an intensity ratio of 3000:1. The projection ratio of the line pattern of the SLM can be varied between 10:1 and 5:1.

The wavelength of the SAW is the ratio of the period of the line pattern on the SLM and the projection ratio $r$: $\lambda_{SAW} = \lambda_{SLM} / r$.

The projection ratio is retrieved in a separate measurement using a silicon substrate. Silicon substrates have no dispersion and the phase velocity of SAW in the 110 direction is exactly known to be 5080. From this velocity and the measured frequency of the SAW propagating through the silicon substrate at different frequencies, the wavelength of the SAW is retrieved and the projection ratio is determined.

The number of periods of the SLM line patterns varies between 20 and 200.

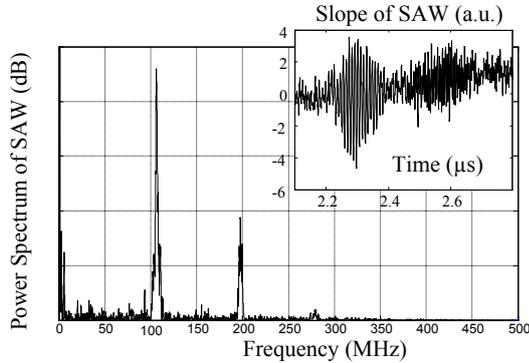

Figure 4: Slope and FFT of the narrow-band SAW from the SLM mask measurement for sample 2 (see Table 2).

## 4. THEORETICAL MODEL

A boundary element method (BEM) model is used to calculate the dispersion of SAWs in the layered system, where the substrate is considered as an infinite anisotropic half-space. A boundary value problem for harmonic waves decaying with depth is deduced from the coupled constitutive equations and from the equation of motion:

$$T_{kl} = c_{klmn} S_{mn} = c_{klmn} \frac{\partial}{\partial x_n} u_m \qquad (4.1)$$

$$\rho \frac{\partial^2 u_i}{\partial t^2} = \frac{\partial T_{ij}}{\partial x_j} \qquad (4.2)$$

where $T_{kl}$ is the stress tensor, $u_i$ the particle displacement, $S_{mn}$ the strain tensor, $c_{klmn}$ the stiffness constants, and $\rho$ the mass density. For each layer the boundary value problem is described by [4]

$$L \cdot r = \lambda \cdot r \qquad (4.3)$$

where $L$ is a matrix operator, $\lambda$ are the eigenvalues and $r$ is a field vector containing both the sources $t_{ij}$ and the wave displacements $u_i$. Since we are only interested in the phase velocity we consider a source having the only one nonzero component $t_{33}$ applied to the surface $x_3 = 0$. The Green's function used relates the wave displacement to the stress source[5]

$$u_3 = \int G_{33}(x-x') t_{33}(x') dx' \qquad (4.4)$$

A Green's function represents the solution of the boundary value problem for an instantaneous point source. It depends on the ratio of the wave number to the angular frequency $k/\omega$, which has the dimension of an inverse velocity and has a pole singularity at the velocity of the surface wave [6].

Rayleigh waves are elliptically polarized in the $x_1$-$x_3$ plane, where $x_1$ is in the propagation direction and $x_3$ is the depth coordinate. In order to identify the phase velocity of the freely propagating Rayleigh wave at a given frequency, the Green's functions of the layered structure are calculated for the source acting along the $x_3$ direction. By constructing the proper Green's function for the layered system, the theoretical dispersion curve is calculated.

By fitting this theoretical model to the measured dispersion curve in a least square sense, a set of elastic and mechanical parameters is deduced, e.g. the Young's modulus, the density and the thickness of the layer.

## 5. RESULTS AND DISCUSSION

### 5.1. Film characterization

*5.1.1. Nanoindenter and weight measurements*
A nanoindenter with a pyramidal indenter has been used to measure the Young's modulus of the SiGe layers. The sample surface was indented at 16 locations and at each location, the Young's modulus and the hardness of the layer has been measured. Only the indentation results at a displacement of less than one tenth of the thickness of the SiGe layer should be evaluated. Otherwise the influence of the substrate has to be considered.

On the other hand, at small indentation displacements of less than 200nm, the results of the 16 indentations diverge too much because of surface roughness.

The Young's modulus of SiGe layers with thicknesses smaller than 1μm ($d_{SiGe} \leq 1\mu m$) could not be analyzed by the nanoindenter method (samples 1A, 2 and 3, see Table 2). A further sample 1B with a thickness of $d_{SiGe} \approx 3.7\mu m$ and a germanium concentration of $c_{Ge} \approx 18\%$ has been indented. The result is a Young's modulus of:

$$E_{SiGe,\ 18\%} = 146 \pm 30 \text{ GPa}$$

SiGe densities were calculated by weighing the wafers after the deposition. In the horizontal batch reactor the wafers, loaded on a quartz wafer boat, were deposited on both sides. After a first weight measurement the SiGe layers were etched back without attacking neither the silicon substrate nor the thermal oxide. For that purpose a dry etching process was used which allows a selective etching of SiGe. After the etching a second weight measurement was performed. The difference of both weights is divided over the volume of the SiGe layer to obtain the mass density. For a number of samples with a germanium concentration of 18% the density was averaged to:

$$\rho_{SiGe,\ 18\%} = 3.04 \pm 0.2 \text{ kg/m}^3$$

*5.1.2. Silicon Oxide $SiO_2$ properties*

The properties of the thermal $SiO_2$ interface layer have been measured separately on an oxidized silicon substrate. After a SAW dispersion measurement, the values of the thickness and the density were entered into the BEM model along with the properties of the silicon substrate and the values of the Young's modulus and poisson ratio were determined. The fitted Young's modulus is in excellent agreement with literature values [7].

| Ellipsometer SE 800 | $d_{SiO2}$ = 2,435 µm ± 2nm |
| Literature [7] | $\rho_{SiO2}$ = 2,2 kg/m³ |
| SAW Measurement | $E_{SiO2}$ = 69,8 Gpa |
| SAW Measurement | $\nu_{SiO2}$ = 0,15 |

Table 1: Properties of the $SiO_2$ interface layer.

*5.1.3. Silicon Germanium properties*

Three SiGe samples with different thicknesses of the SiGe layer $d_{SiGe}$ and germanium concentrations $c_{ge}$ have been analyzed. The SiGe layers were structured using plasma etching with $SF_6$. The generated steps heights were measured using a profilometer. The chemical composition of the SiGe layers was investigated by X-ray photoelectron spectroscopy (XPS).

|  | 1A | 2 | 3 |
|---|---|---|---|
| $d_{SiGe}$ | 1000 ± 50 nm | 700 ± 50 nm | 900 ± 50 nm |
| $c_{ge}$ | ~ 18% | ~ 60% | ~ 40% |
| $d_{SiO2}$ | 2435 ± 2nm | 2435 ± 2nm | no $SiO_2$ |

Table 2: Properties of the layer systems.

To gain more insight into the film morphology of the SiGe layers several SEM cross sections were analysed. Figure 5a shows the 1µm thick SiGe layer of sample 1A. The low surface roughness (~20 nm) is one of the requirements for SAW film characterisation. Thicker films with increased grain sizes consequentially enhance the surface roughness up to 40 nm as can be seen in the SEM cross section in Figure 5b. Sample 2 and sample 3 are processed with comparable reduced surface roughnesses as Sample 1A.

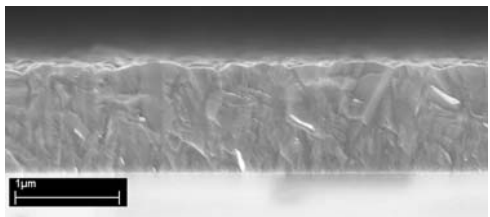

Figure 5.a: Cross section SEM of 1µm thick SiGe (Sample 1A).

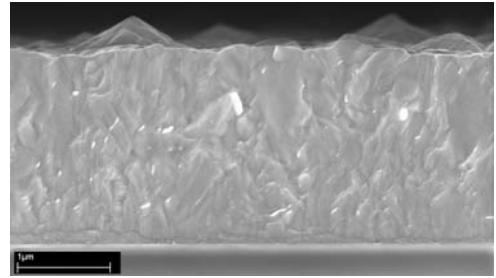

Figure 5.b: Cross section SEM of 2µm thick SiGe (Similar LPCVD-parameters as sample 1A).

For investigating the crystallinity of the SiGe layers, X-ray diffraction (XRD) provides a useful characterization method. The SiGe layers analyzed in this report indicate no appreciable texture. The graph in Figure 6 shows the XRD spectrum of the sample 2 with about 60% Germanium and a thickness of 700 nm.

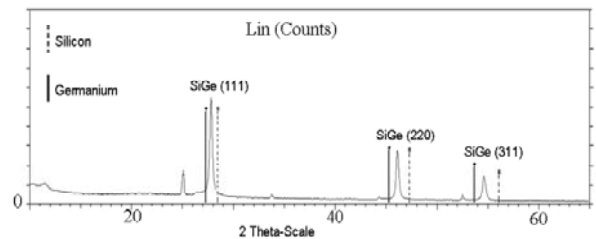

Figure 6: XRD spectrum of sample 2.

Comparing the indicated SiGe peaks no significant crystal direction is dominating. As can be seen in the XRD-spectrum the ratio of the detected crystal planes is balanced and no distinct texture is predominating. In contrast to these homogeneous SiGe films increasing layer thicknesses result in a textured material. The LPCVD film growth characteristics can be considered as a reason for the arising texture over film thickness. Starting from a small-grained nucleation zone columnar crystallites with increasing grain sizes and preferred growing directions are predominating in thicker films. With the objective to determine the properties of SiGe without a distinct texture, the second reason for the use of thin SiGe layers in the SAW experiments follows. The influence of the change in crystal directions on the SAW measurements can be completely eliminated in thin layers. In the case of thicker layers, SAW can be used to detect the change in the Young's Modulus over the thickness of textured SiGe films including such distinct textures if desired.

### 5.2. Excitation of SAWs with glass masks

In Figure 7, the measured dispersion curves representing the SAW phase velocities for the different mask periods are plotted in dots. The line curves represent the fit of the BEM model to the measured dispersion curves. The result of this fit is a set of physical parameters (Table 3).

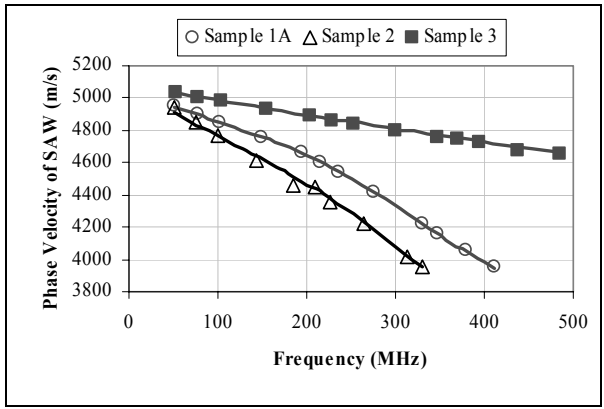

Figure 7: Dispersion curves for SiGe samples.

For the samples 1A, 2 and 3, a linear relation between the Young's modulus and the density on one hand and the germanium concentration on the other hand has been assumed:

$$E_{SiGe} = E_{Si} - c_{Ge} \cdot (E_{Si} - E_{Ge}) \quad (5.1)$$
$$\rho_{SiGe} = \rho_{Si} + c_{Ge} \cdot (\rho_{Ge} - \rho_{Si}) \quad (5.2)$$

For the sample 3, because of the linear behavior of the dispersion curve, only the germanium concentration could be fitted. For this sample the thickness has been set to the fixed value of 0.9 μm.

|  |  | 1A | 2 | 3 |
|---|---|---|---|---|
| SAW | $c_{Ge}$ (%) | **17.9** | **62.4** | **41.6** |
|  | $E$ (GPa) | 155 | 142 | 148.36 |
|  | $\rho$ (kg/m³) | 2.86 | 4.19 | 3.57 |
|  | $d_{SiGe}$ (μm) | **1.02** | **0.71** | not fitted |
| XPS | $c_{Ge}$ (%) | 18 | 60 | 40 |
| (5.1) | $E$ (Gpa) | 155 | 143.2 | 148.8 |
| (5.2) | $\rho$ (kg/m³) | 2.87 | 4.24 | 3.53 |
| Profilometer | $d_{SiGe}$ (μm) | 1 | 0.7 | 0.9 |

Table 3: The fit results for SiGe samples measured with SAW masks are printed in bold. Values from the XPS measurement and the equations (5.1) and (5.2) are also given.

### 5.3. Excitation of SAWs with an SLM mask

In Figure 8 the dispersion measurements with a glass mask and an SLM generated SAW for the sample 1A are compared. The number of periods from the SLM generation is relatively small: Between 20 and 200 for a frequency range between 50 and 500MHz. Therefore, the FFT peaks are wider at low frequencies and the velocities retrieved from these peaks differ by as much as 5% from those from the mask excitation. On the other hand, the dispersion curve from the SLM excitation remains systematically below the dispersion curve from the mask excitation. We consider that the main reason for this systematic error relies in the inaccurate determination of the projection ratio of the optical setup. This ratio was determined to be at

$$r = 9.1 \pm 0.23$$

The fit result for the dispersion curve with SLM generated SAWs is a germanium concentration of 29.6%, which is higher than the 18 % from the measurements in Table 3.

At a projection ratio of 8.9, the dispersion curve from the SLM mask generated SAW and the dispersion curve from the glass mask measurements overlap. At this projection ratio, the germanium concentration is fitted to 16.1%.

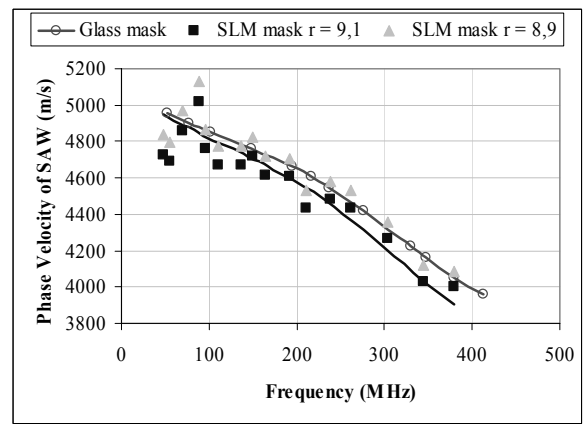

Figure 8: Dispersion curves from the glass mask and the SLM excitation.

### 6. CONCLUSION

The results obtained for the Young's modulus and the density of the three SiGe samples with different germanium concentrations confirm the equations 5.1 and 5.2. It is sufficient to know the germanium concentration of the SiGe layer to obtain the Young's modulus and the density.

The Young's modulus and the density obtained for the germanium concentration of ~18% (samples 1A and 1B) is in excellent agreement with the results of the nanoindenter and the weight measurements. An excellent agreement was also found for the thicknesses of the samples 1A and 2 compared with the profilometer measurements.

The potential of the excitation of SAW with an LCD mask is demonstrated. Here the challenge relies in knowing the projection ratio with an accuracy better than 1%.

## 7. ACKNOWLEDGMENTS

The authors thank F. Schatz, M. Lammer and H. Eisenschmid from Robert Bosch GmbH, as well as W. Dötzel from the Technical University of Chemnitz, for helpful discussions.